\documentclass[twocolumn,reprint,showpacs,superscriptaddress,nofootinbib,10pt]{revtex4-1} 
\usepackage{amsmath,amssymb,graphicx}
\usepackage{placeins}
\usepackage[none]{hyphenat}
\usepackage[compact]{titlesec}
\usepackage[usenames,dvipsnames]{color}
\usepackage[hyperfootnotes=false]{hyperref}
\usepackage[hyperfootnotes=false]{hyperref}
\hypersetup{colorlinks=true,citecolor=Blue,linkcolor=Blue,urlcolor=Blue}
\begin{document}

\title{\large{Supersymmetry generated one-way invisible $\mathcal{PT}$-symmetric optical crystals}}

\author{Bikashkali Midya}
\email{bikash.midya@gmail.com}
\affiliation{Physique Nucl\'eaire et Physique Quantique, Universit\'e libre de Bruxelles, B-1050 Brussels, Belgium. \vspace{.2 cm}}

\begin{abstract}
{ We use supersymmetry transformations to design transparent and one-way reflectionless (thus unidirectionally invisible) complex optical crystals with balanced gain and loss profiles. The scattering co-efficients are investigated using the transfer matrix approach. It is shown that the amount of reflection from the left can be made arbitrarily close to zero whereas the reflection from the right is enhanced arbitrarily (or vice versa).} \vspace{.1 cm}

\end{abstract}

\pacs{11.30.Er, 42.25.Bs, 03.65.Nk, 11.30.Pb \vspace{.1 cm}}
\maketitle 

 We see an object because light bounces off it. If this scattering of light could be cloaked and if the object does not absorb any light then it would become invisible. Although {\it invisibility} has been a subject of science fiction for millennia, the recent discovery of metamaterials is opening up the possibility of practical demonstrations of cloaking devices \cite{Le06,PSS06,CCS10,Va+09}. A properly
designed metamaterial shell surrounded around a given object
can drastically conceal its scattering for any angle of incidence, making it almost undetectable. Different techniques like the coordinate transformation technique\cite{Le06}, and the scattering cancellation technique \cite{AE05}, are suggested to design cloaking from electromagnetic waves. The realization of a coordinate transformation cloak, which is able to hide a copper cylinder at microwave frequency, has been recently reported \cite{Sc06}. The concept of cloaking has also been extended to the
quantum and acoustic domains, realizing matter-wave \cite{Zh+08,FA13}
and acoustic cloaks \cite{FEGM08,ZXF11}. Nevertheless, cloaking in visible light, hiding more complex shapes and materials, still remains distant.

Very recently, it has been discovered \cite{Ma+08,Ru+10,Be08,Zh+10,Gu+09,Lo10R,CGS11,Sc10,Ca+13,Zh+13} that light propagation can also be influenced substantially by controlling the parity-time ($\mathcal{PT}$) symmetry in such a way that amplification and loss balance each other. Most interestingly, as opposed to wrapping a scatterer with a cloak,  $\mathcal{PT}$-symmetric material can become one-way invisible as a result of  spontaneous $\mathcal{PT}$-symmetry breaking. Such unidirectional invisibility has been predicted \cite{Li+11} by Bragg scattering in sinusoidal complex crystal of finite length : $\Delta n(z) = b (\cos 2\pi z/a + i \sigma \sin 2\pi z/a)$ near its symmetry breaking point $\sigma =1$. A ray of light when it hits one side of such a material is transmitted completely without any reflection. In this same regime the transmission phase also vanishes, which is compulsory for avoiding detectability. When the transmittance and (left, right) reflectance are analytically expressed \cite{Lo11,Jo12} in terms of the modified Bessel functions, it becomes clear on closer inspection that there is, however, a very small deviation of left reflectance from $0$ (varies rapidly on the scale of $10^{-6}$ for $b=0.001$). The transmission is also not perfect in amplitude or phase.  Moreover, the unidirectional invisibility is ambiguous for a crystal with length $L > 2 \pi^3 / b^2 a^3$ \cite{Lo11}. Thus, at the $\mathcal{PT}$-symmetry breaking point the sinusoidal crystal appears to be one-way invisible solely for a shallow grating which indeed is realized by recent experiments on a $\mathcal{PT}$-synthetic photonic lattice \cite{Re+12,YZ13}.

On the other hand, {\it nonrelativistic} supersymmetry (SUSY) transformations are shown \cite{CW94,Mi+13,MHC13,LV13,LD13,BC06,Fe14} to be useful in the framework of optics to synthesize new optical structures. In particular, SUSY has provided a method to generate an optical medium with defects that can not be detected by an outside observer \cite{LD13}, to obtain transparent interface separating two isospectral but different crystals \cite{LV13}, and to create a family of isospectral potentials to optimize quantum cascade lasers \cite{BC06}. In ref.\cite{MHC13}, SUSY has been used to generate a complex optical potential with real spectra, even their shape violate $\mathcal{PT}$-symmetry. Further, SUSY photonic lattices \cite{Mi+13} are used
to design lossless integrated mode filtering arrangements.

Our purpose here is to use SUSY transformations of the sinusoidal complex crystal at its symmetry breaking point to design one-way invisible crystals with sophisticated shape and structure. The scattering co-efficients for these crystals are investigated using the transfer matrix approach \cite{Mo09,Mo13,GCS12}. Precisely, we have derived the relationship between the transfer matrices of the initial crystal $V^{(0)}$ and its $n$th order isospectral crystals $V^{(n)}$. This reveals that the corresponding transmission coefficients do not alter their values, whereas the values of left and right reflection coefficients do. The left (right) reflectivity can be diminished (enhanced) arbitrarily using higher order SUSY transformations. For instance the magnitude of left reflection (for $b=0.001$) is reducible from $10^{-6}$ to $10^{-10}$ after two transformations.

\begin{figure}[h!]
     \includegraphics[width=4.15cm,height=3.3cm]{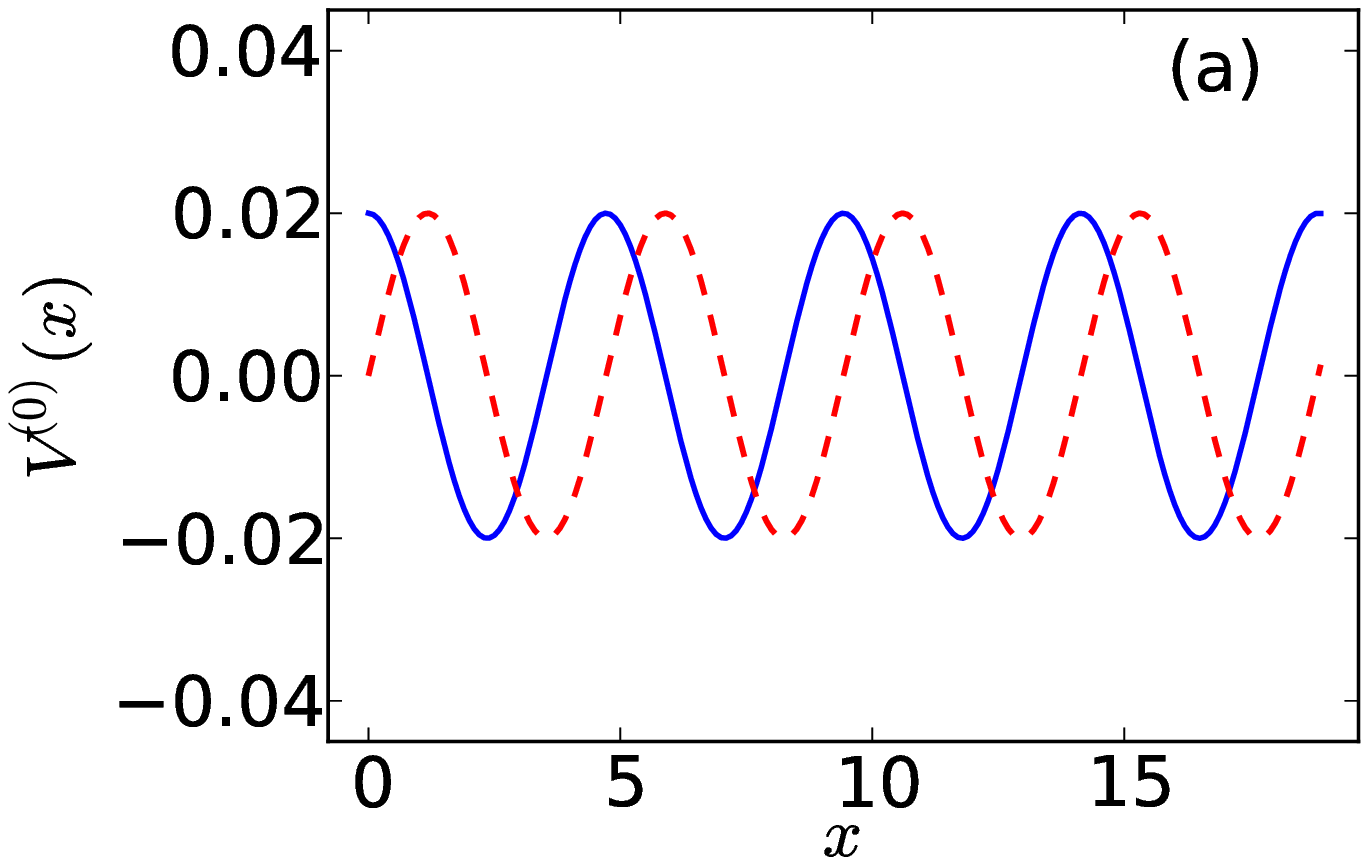}  \includegraphics[width=4.15cm,height=3.3cm]{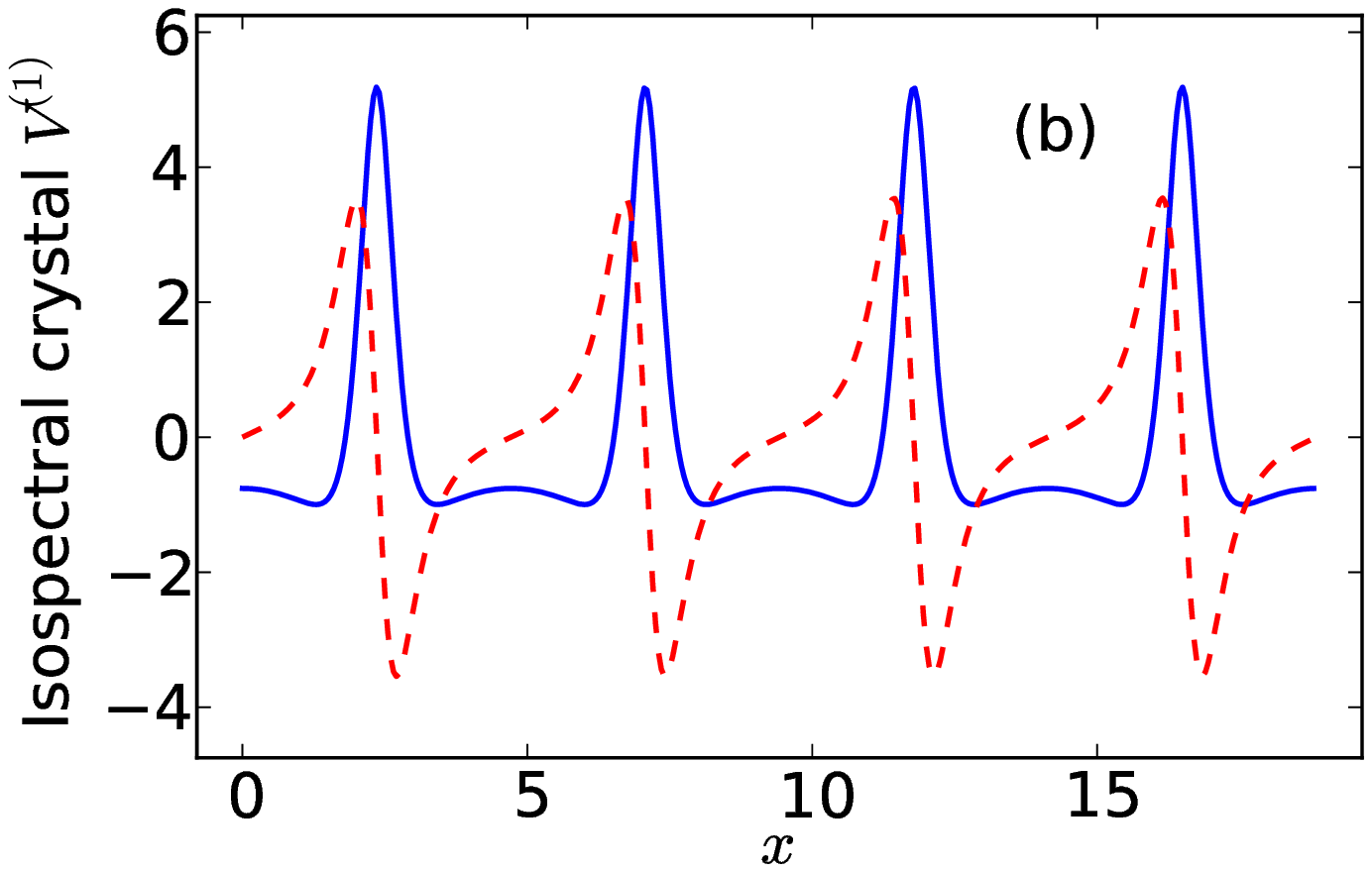}
        \includegraphics[width=4.15cm,height=3.3cm]{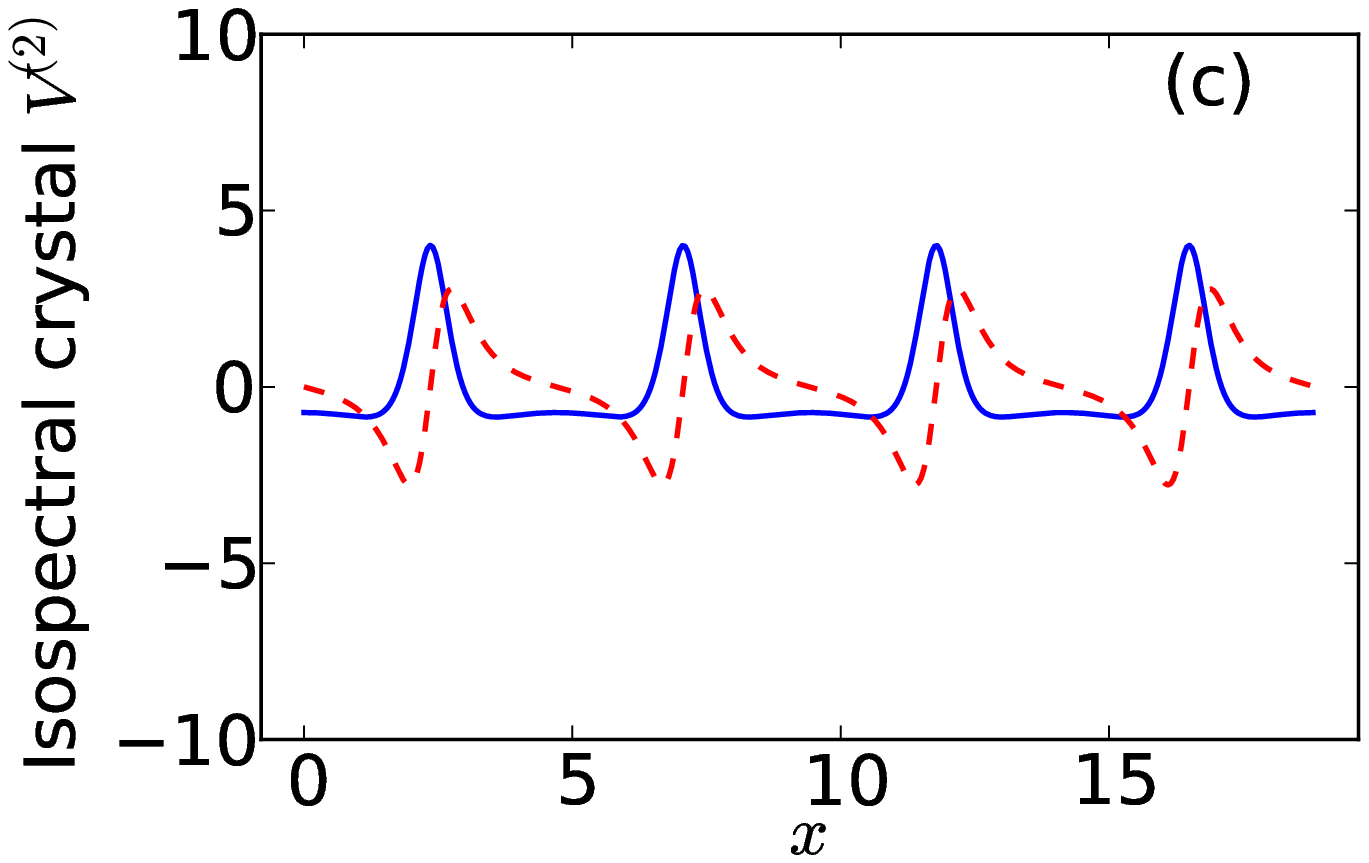}  \includegraphics[width=4.15cm,height=3.3cm]{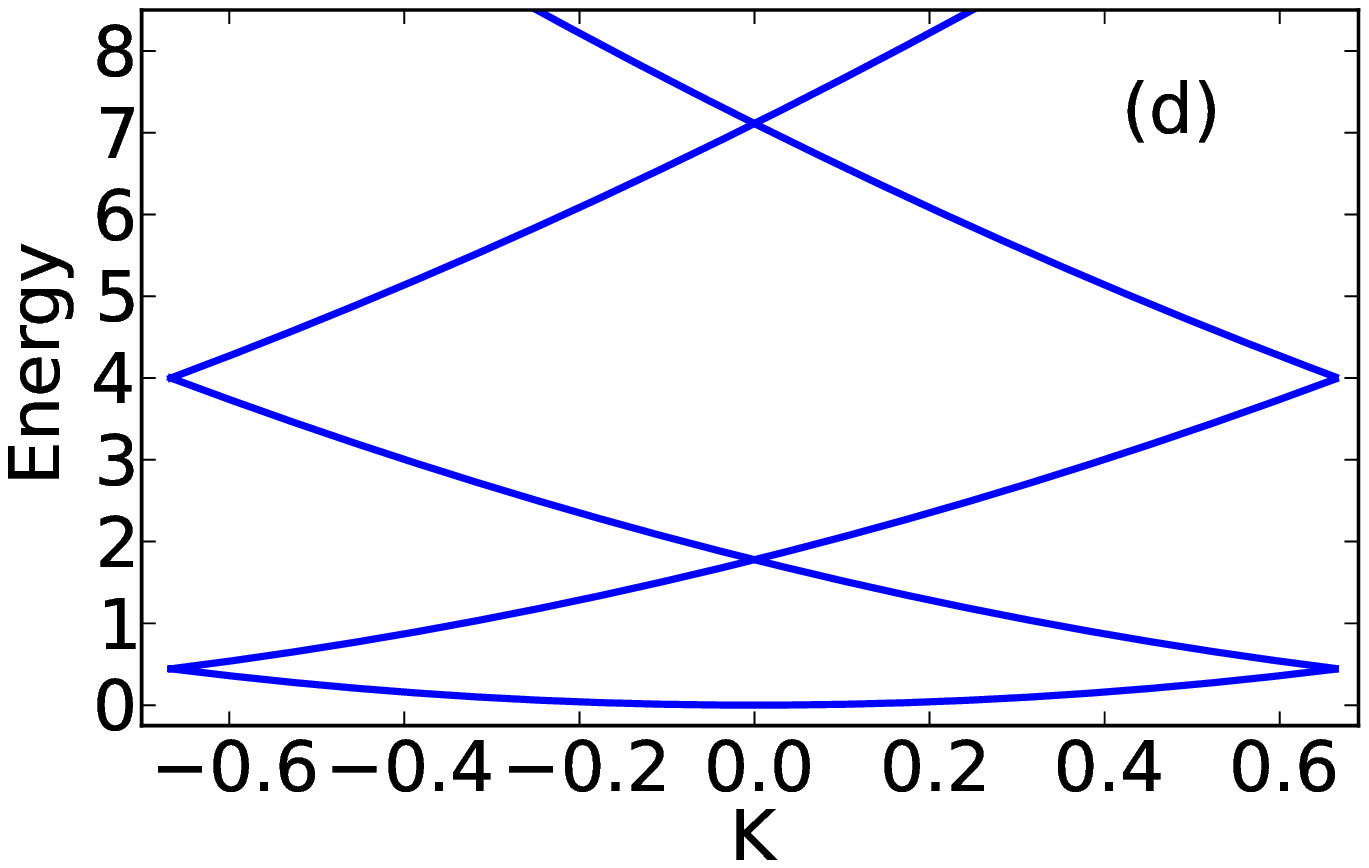}
      \caption{(Color online) Real (solid line) and imaginary (dashed line) parts of (a) the crystal given in (\ref{e20}), (b) isospectral crystal $V^{(1)}$ obtained using first-order SUSY transformation, (c) isospectral crystal $V^{(2)}$ obtained using second-order SUSY transformations. (d) Reduced zone energy band structure. Here we have considered $b = 0.02, a= 3\pi/2, \xi_0 =0.44$ and $\xi_1 =0.9$.}\label{f1}
      \end{figure}

{\it Construction of isospectral crystals:} We consider a $\mathcal{PT}$-symmetric relative dielectric constant $n(z) = n_0^2 [1+ \Delta n(z)]$, where $\Delta n(z+a) = \Delta n(z)$ is the complex refractive index whose imaginary part represents either gain or loss. The variation in $n(z)$ is measured along the longitudinal $z$ direction in $(0,L)$. In this setting, a time-harmonic electric field of frequency $\omega$ obeys the scalar Helmholtz equation \cite{Li+11,Lo11}, which is formally identical to the time-independent Schr\"odinger equation for the wave function $\psi$: 
\begin{equation}
H \psi(z) = -\frac{d^2\psi(z)}{dz^2} - V(z) \psi(z) = E \psi(z), \label{e1}
\end{equation}
provided $V \propto \Delta n$ and $\omega$ is very close to the Bragg frequency $\omega_B = c \pi/(n_0 a).$ The $\mathcal{PT}$-symmetry of the refractive index translates into the potential such that $V(L-z)^* = V(z).$  It is the following potential that we wish to consider here as a reference potential to construct isospectral crystals:
\begin{equation}
V(z) = b e^{2 i \pi z/a}, \quad \quad 0< z <L.\label{e20}
\end{equation}
In addition, later in the context of scattering we will consider $V(z) = ~ constant$ for $z<0$ and $z>L$. The spectral problem for this potential is well-studied in \cite{Ga80,CJT98,MRR10,GJ11}. The spectrum of $H$ is the semi-infinite real axis and there is no band gap. The equation (\ref{e1}) can readily be reduced to the Bessel modified differential equation after changing of variable $y(z) = a\sqrt{b}/\pi \exp(i \pi z/a)$
\begin{equation}
y^2 \frac{d^2 \psi}{dy^2} + y \frac{d\psi}{dy} - (y^2 + q^2) \psi = 0,
\end{equation}
where $q = a\sqrt{E}/\pi$. Hence for a non integer $q$ the most general solutions can be written as 
\begin{equation}
\psi(z,E) = \alpha_1 I_{\frac{a\sqrt{E}}{\pi}}(y) + \alpha_2 I_{-\frac{a\sqrt{E}}{\pi}}(y), \label{e15}
\end{equation}
where $I_q(y)$ is the modified Bessel function of the first kind. Moreover for the crystal momentum $k = \sqrt{E}$ we have $\psi(z+a) = e^{i k a} \psi(z)$. This implies that $\psi(z)$ is a Bloch wave function. The potential and corresponding reduced zone band structure are shown in figures \ref{f1}(a) and \ref{f1}(d), respectively. 

Now, we consider two periodic potentials $V^{(0)}$ $(=V)$ and $V^{(1)}$ describing two crystals with different unit cells but with the same lattice period, i.e. $V^{(0)}(z) \ne V^{(1)}(z)$ but $V^{(0),(1)}(z+a) = V^{(0),(1)}(z)$. Then these two crystals are said to be isospectral if they have an identical energy band structure. Like the Hermitian case \cite{CKS02,DF98}, SUSY transformations also enable \cite{LD13} one to easily construct a new complex periodic potential $V^{(1)}$, which is isospectral\footnote{Here we will not consider the self-isospectral crystals (which are such that $V^{(0)}$ and $V^{(1)}$ are related by simple translation, $z\rightarrow z+ \alpha$ or inversion $z\rightarrow -z$)} to $V^{(0)}$. To this aim, the Hamiltonian $H^{(0)}$ is written in factored form $H^{(0)} = B_1 A_1 + \xi_0$ with the help of following two first order linear operators
\begin{equation}
A_1 = -\frac{d}{dz} + w_1(z), ~~ B_1 = \frac{d}{dz} + w_1(z),\label{e10}
\end{equation}  
where $\xi_0$ is the energy of factorization and the superpotential $w_1(z)$ is defined in terms of Bloch solution:
\begin{equation}
w_1(z) = u_0'(z) u_0^{-1}(z), ~~~ H^{(0)} u_0(z) = \xi_0 u_0(z).
\end{equation} 
Consequently, $V^{(0)}$ can be expressed as $V^{(0)} = -(w_1^2 + w_1')$. Note here that for a $\mathcal{PT}$-symmetric complex $V^{(0)}$, $w_1$ is complex and $B_1 \ne A_1^\dagger$. To $H^{(0)}$ there corresponds the partner Hamiltonian $H^{(1)} = A_1 B_1 + \xi_0$ with 
\begin{equation}
V^{(1)}(z) = -(w_1^2 - w_1') = V^{(0)}(z) + 2w_1'(z).\label{e22}
\end{equation}
For a periodic and exact $\mathcal{PT}$-symmetric potential $V^{(0)}$, the  Bloch wave function $u_0$ implies that the superpotential $w_1$ is periodic and anti-$\mathcal{PT}$-symmetric i.e. $w_1(z+a) = w_1(z)$ and $w_1(L-z)^* = - w_1(z)$. Hence Eq. (\ref{e22}) clearly shows that $V^{(1)}$ respects the condition of periodicity and $\mathcal{PT}$-symmetry. For periodic systems, the two zero modes (which are the solutions of $A_1 \psi_0^{(0)} = 0$ and $B_1 \psi_0^{(1)} = 0$, respectively)
\begin{equation}
\psi^{(0),(1)}_0(z) = e^{\pm\int^z w_1(t) dt}\label{e100}
\end{equation}
will belong to the Hilbert space if they satisfy the Bloch condition mentioned earlier. Now using the periodicity condition $w_1(z+a) = w_1(z)$ in equation (\ref{e100}) we have 
\begin{equation}
\psi_0^{(0),(1)}(z+a) = e^{\pm \phi_a} \psi_0^{(0),(1)}(z), ~ ~~ \phi_a = \int_0^{a} w_1(t) dt.\label{e101}
\end{equation}
Clearly, $\psi_0^{(0),(1)}$ will be Bloch wave functions if $\pm\phi_a = i k a$. In other words for periodic case the SUSY is said to be unbroken if $Re(\phi_a) = 0$, otherwise it is broken. Consequently, the two zero modes $\psi_0^{(0),(1)}$ either both are Bloch functions (in which case SUSY is unbroken), or neither of them are Bloch functions (when $Re(\phi_a) \ne 0$ and SUSY is broken).  
Thus in the periodic case, irrespective of whether SUSY is broken or unbroken, the potentials $V^{(0),(1)}$ are always strictly isospectral. For $\mathcal{PT}$-symmetric complex potentials, the superpotential is anti-$\mathcal{PT}$-symmetric i.e. the real part of $w_1(z)$ is an odd function. This implies that the real part of the integral in Eq.(\ref{e101}) is always equal to zero. Thus for $\mathcal{PT}$-symmetric periodic systems the SUSY is always unbroken and the energy spectra are strictly identical. Supersymmetry also allows one to connect the solutions of $H^{(1)}$ to those of $H^{(0)}$ via the relation $\psi^{(1)}(z,E) = A_1 \psi^{(0)}(z,E)$. Consequently, if $\psi^{(0)}$ is a Bloch wave function, then so is $\psi^{(1)}$.

The above technique can be applied to obtain another new Hamiltonian $H^{(2)} = -d^2/dz^2 - V^{(2)}$ isospectral to $H^{(1)}$ such that $H^{(1)} = B_2 A_2 + \xi_1$ and $H^{(2)} = A_2 B_2 + \xi_1$.  Here the operators $A_2, B_2$ has the same form as in equation (\ref{e10}) but with different superpotential $w_2(z) = v_1'(z) v_1^{-1}(z)$, where the factorization function $v_1$ and energy of factorization $\xi_1$ $(\ne \xi_0)$ satisfy $H^{(1)} v_1(z) = \xi_1 v_1(z)$.  The solutions of $H^{(2)}$ with
\begin{equation}
V^{(2)} = V^{(1)} + 2w_2' = V^{(0)} + 2(w_1 + w_2)'
\end{equation} 
are given by $\psi^{(2)}(z,E) = A_2\psi^{(1)}(z,E) = A_2 A_1 \psi^{(0)}(z,E)$. Repeating the procedure $n$ times one gets
\begin{equation}
V^{(n)} = V^{(0)} + 2 \mathcal{W}_n', ~~ \mbox{where} ~~ \mathcal{W}_n = \sum\limits_{k=1}^{n} w_k,\label{e7}
\end{equation}
and $w_k = v'_{k-1} v_{k-1}^{-1}$, $v_{k-1}$ (with $v_0 = u_0$) being the solution of $H^{(k-1)}$ at the factorization energy $\xi_{k-1}$. The chain of these $n$ SUSY transformations allows one to find the solution of the new Hamiltonian 
$H^{(n)} = -d^2/dz^2 - V^{(n)}$, in the following form 
\begin{equation}
\psi^{(n)}(z,E) = A_n A_{n-1}.... A_1 ~\psi^{(0)}(z,E). \label{e5}
\end{equation}
      \FloatBarrier
Though the above mentioned formulas are very compact and elegant, they are not very convenient for practical applications because in order to obtain $V^{(n)}$ from $V^{(0)}$ one has to consider all the $(n-1)$ intermediate steps. This difficulty can be overcome by writing the expressions another way in terms of the solutions of the initial reference Hamiltonian only. Since all the solutions $v(z,\xi_k)$ of $H^{(k)}$ are related to the solution $u(z,\xi_k)$ of $H^{(0)}$ by the relation (\ref{e5}), after a cumbersome but straight forward calculations we can rewrite 
\begin{subequations}
\begin{equation}
 \mathcal{W}_n = \frac{W[u_0,...,u_{n-1}]'}{W[u_0,...,u_{n-1}]}, \quad \quad  n=1,2...\label{e6a}
 \end{equation}
 \begin{equation}
  V^{(n)} = V^{(0)} + 2 \left(\frac{W[u_0,...,u_{n-1}]'}{W[u_0,...,u_{n-1}]}\right)'\label{e6b}
 \end{equation}
 \begin{equation}
 \psi^{(n)} = W[u_0,...,u_{n-1},\psi^{(0)}] ~ \left(W[u_0,...,u_{n-1}]\right)^{-1},
\end{equation}
\end{subequations}
where $W[ ~]$ is the Wronskian determinant with $W[u_0] = u_0$ so that $\mathcal{W}_1 = w_1$. Such expressions are known as Crum-Krein formulas \cite{Cr55,Kr57}. 

 Now using the equation (\ref{e6b}) and taking $u_k = u(z,\xi_k)$ as the (Bloch) factorization solutions given in equation (\ref{e15}), it is not difficult to construct the crystals isospectral to the one given in (\ref{e20}). Specifically, for $\alpha_1=0$ and $\alpha_2 =1$, the first order isospectral crystal reduces to the following simple functional form  
\begin{equation}
V^{(1)} = b e^{2 i \pi z/a} + 2 \partial_{zz} \ln I_{\frac{-a\sqrt{\xi_0}}{\pi}}\left(\frac{a\sqrt{b}}{\pi} e^{i \pi z/a}\right).
\end{equation} 
For illustration, we have plotted the real and imaginary parts of the two isospectral crystals $V^{(1),(2)}$ in figure \ref{f1}(b), and \ref{f1}(c), respectively. These two crystals have the same energy band structure as shown in figure \ref{f1}(d).  Here we have considered the two factorization functions as $u_0 = I_{-q_0}(y)$ and $u_1 = I_{q_1}(y)$ at the energies $\xi_0=0.44$ and $\xi_1=0.9$, respectively.

{\textit {Transfer matrix \& unidirectional invisibility}}: In general, for a localized potential $V(z)$, restricted to the interval $0< z <L$, the scattering solution can be written as 
   \begin{equation}
   \Psi(z) = \left\{\begin{array}{lll}
   \beta_{l\rightarrow} ~e^{i p z} +\beta_{l\leftarrow} ~e^{- i p z}, ~ z <0\\
   \psi(z) , ~~~~ 0< z <L\\
    \beta_{r\rightarrow} ~ e^{i p (z-L)} + \beta_{r\leftarrow} ~e^{- i p (z-L)}, ~ L < z
   \end{array}\right.\label{e25}
   \end{equation}
where `$\rightarrow$' and `$\leftarrow$' denote the forward and backward direction of the wave propagation, respectively; $l, r$ denote the left-hand $(z<0)$ and right-hand $(z >L)$ side of the crystal, respectively. To find the scattering amplitudes, one has to first solve Eq. (\ref{e1}) for $\psi(z)$ in $(0,L)$. Then, invoking the appropriate boundary conditions at
$0$ and $L$ (typically, continuity of $\psi(z)$ and it's derivative) one
obtains two linear equations among the coefficients. These can be solved for the two right-side amplitudes in terms
of the other two, and the result can be expressed as a matrix
equation:
\begin{equation*}
\left(\begin{array}{c}
\beta_{r\rightarrow} \\ 
\beta_{r\leftarrow}
\end{array}\right) = \mathcal{M}(p) \left[\begin{array}{c}
 \beta_{l\rightarrow}\\\beta_{l\leftarrow} 
\end{array} \right], ~~~ \mathcal{M} = \left[\begin{array}{cc}
 M_{11} & M_{12} \\ 
 M_{21} & M_{22}
\end{array}  \right].
\end{equation*}   
This $2\times2$ transfer matrix $\mathcal{M}$ is unimodular (det $\mathcal{M} = 1$) and the elements are related to the transmission $(t)$ and reflection $(r)$ coefficients for left-side $l$ and right-side $r$ incidence by $t_l = t_r = t = M_{22}^{-1}, r_l = - M_{21} M_{22}^{-1}$ and $r_r = M_{12} M_{22}^{-1}$.
Some distinctive features of the $\mathcal{PT}$-symmetric scattering (as discussed in \cite{GCS12,CDV07}) are as follows: the transmission coefficient does not depend on the incidence side like in a Hermitian optical crystal. The left and right reflection coefficients are in general unequal $|r_l| \ne |r_r|$. Moreover the two reflectances $\mathbb{R}_{l,r} = |r_{l,r}|^2$ and the transmittance $\mathbb{T} = |t|^2$ do not add up to unity (i.e. $ \mathbb{R}_{l,r}+\mathbb{T} \ne 1$), instead they satisfy the generalized unitarity relation 
\begin{equation}
|\mathbb{T}-1| = \sqrt{\mathbb{R}_l \mathbb{R}_r}. \label{e102} 
\end{equation}
Thus in the $\mathcal{PT}$-symmetric non-Hermitian case, the geometric mean of the two reflectances, $\sqrt{\mathbb{R}_l \mathbb{R}_r}$, replaces the single reflectance $\mathbb{R}$ in the conventional flux conserving relation for the Hermitian system. A $\mathcal{PT}$-symmetric potential is said to be invisible from the left (right) if $\mathbb{R}_l = 0$ ($\mathbb{R}_r =0$) together with $\mathbb{T} = 1.$ 

 If we indicate by $\mathcal{Z}$ as the fundamental matrix \cite{Lo11} of equation (\ref{e1}), which relates the values of $\psi(z)$ and $\psi'(z)$ at $z=0$ and $z=L$, then 
 \begin{equation}
 \mathcal{Z} = \left[\begin{array}{cc}
  \psi_1(L) & \psi_2(L) \\ 
  \psi_1'(L) & \psi'_2(L)
  \end{array} \right] \times \left[\begin{array}{cc}
    \psi_1(0) & \psi_2(0) \\ 
    \psi_1'(0) & \psi'_2(0)
    \end{array} \right]^{-1} \label{e16}
 \end{equation}
 where $\psi_{1,2}(z)$ are the two linearly independent solutions of Eq. (\ref{e1}). The transfer matrix is related to the fundamental matrix by the following relation
 \begin{equation}
 \mathcal{M} = T^{-1} \mathcal{Z}(p) T, ~~~~~ T = \left[\begin{array}{cc}
 1 & 1 \\ 
 i p & -ip
 \end{array} \right].
 \end{equation}
  
In the following we derive the relationship between the transfer matrices associated with a given initial crystal and its $n$-th order isospectral partner. It is important to mention here that in order to guarantee that the scattering to take place in both the SUSY periodic potentials, it is necessary that the potentials $V^{(0),(n)}$ are asymptotically constant in the region $z \in (-\infty,0]\cup [L,\infty)$. Without loss of generality we assume the constant to be $w_1^2(0)$ (where we have set $\mathcal{W}'_n(z) \rightarrow 0$ in the same region and assumed $w_1(0) = w_1(L)$). Hence the momentum appearing in equation (\ref{e25}) is given by $p = |\sqrt{E-w_1^2(0)}|$.\\

\noindent {\bf Theorem:} If $\mathcal{M}_{0}$ and $\mathcal{M}_{n}$ are the corresponding transfer matrices of $V^{(0)}$ and $V^{(n)}$, respectively, then for $n=1,2,3,...$
\begin{equation}
\mathcal{M}_{n} = \mathcal{D}_n^{-1} \mathcal{M}_{0} \mathcal{D}_n, ~~ \mathcal{D}_n = \left[\begin{array}{cc}
\prod\limits_{k=1}^{n}\frac{i}{p + i w_k(0)} &  0\\ 
0 & \prod\limits_{k=1}^{n} \frac{(-i)}{p - i w_k(0)}
\end{array}\right].
\end{equation}
  
{\bf Proof:} Let us first consider the case with $n=1$.  In this case recalling the relationship $\psi^{(1)} = A_1\psi^{(0)}$ we have  
{\small \begin{equation*}
\textstyle  \left[\begin{array}{cc}
 \psi_1^{(1)}(z) & \psi^{(1)}_2(z) \\ 
 {\psi_1^{(1)}}'(z) & {\psi_2^{(1)}}'(z)
 \end{array} \right] = \left[\begin{array}{cc}
  w_1 & -1 \\ 
  E-w_1^2 & w_1
 \end{array}  \right] \left[\begin{array}{cc}
  \psi_1^{(0)}(z) & \psi_2^{(0)}(z) \\ 
  \psi_1^{(0)'}(z) & \psi_2^{(0)'}(z)
  \end{array} \right]
 \end{equation*}}
If $\mathcal{Z}_0$ is the fundamental matrix for $V^{(0)}$ then the above relation suggests that the fundamental matrix $\mathcal{Z}_1$ for the potential $V^{(1)}$ can be expressed as
 \begin{equation}
 \mathcal{Z}_{1} = \mathcal{B}_1 \mathcal{Z}_{0} \mathcal{B}_1^{-1}, ~~~~ \mbox{where} ~~ \mathcal{B}_1  =  \left[\begin{array}{cc}
  w_1(0) & -1 \\ 
   p^2 & w_1(0)
  \end{array}  \right].
 \end{equation} 
Here we have used $w_1(0) = w_1(L)$. Hence the corresponding transfer matrix $\mathcal{M}_1$ for $V^{(1)}$ is reduced to
\begin{equation}
\mathcal{M}_1 = T^{-1} \mathcal{Z}_1 T = \mathcal{D}_1^{-1} \mathcal{M}_0 \mathcal{D}_1,
\end{equation}
where $\mathcal{D}_1 = T^{-1} \mathcal{B}_1^{-1} T$ is a diagonal matrix with non-vanishing entries $i/[p+iw_1(0)]$ and $-i/[p-iw_1(0)]$. Iterating the above procedure $n$ times and using equation (\ref{e5}) we have in general 
\begin{equation}
\mathcal{Z}_n = \left(\mathcal{B}_n \mathcal{B}_{n-1} \dots \mathcal{B}_1\right) ~ \mathcal{Z}_0 ~\left(\mathcal{B}_n \mathcal{B}_{n-1} \dots \mathcal{B}_1\right)^{-1}
\end{equation}
and hence $\mathcal{M}_n = \mathcal{D}_n^{-1} \mathcal{M}_0 \mathcal{D}_n$, where 
\begin{equation*}
\mathcal{B}_k = \left[\begin{array}{cc}
  w_k(0) & -1 \\ 
   p^2 & w_k(0)
  \end{array}  \right] ~~ \mbox{and} ~~ \mathcal{D}_n = T^{-1} \textstyle \prod\limits_{k=1}^{n} \mathcal{B}_k^{-1} T  
\end{equation*}
Explicit calculation reveals that $\mathcal{D}_n$ is a diagonal matrix with diagonal entries $\prod\limits_{k=1}^{n} \frac{i}{p + i w_k(0)}$ and  
$ \prod\limits_{k=1}^{n} \frac{(-i)}{p - i w_k(0)}$. 

An immediate consequence of the above theorem is that the transmittance and reflectance of the two isospectral crystals $V^{(0)}$ and $V^{(n)}$ are related by 
\begin{equation}
t^{(n)} = t^{(0)}, ~~~~ r_{l,r}^{(n)} = (-1)^n \textstyle \prod\limits_{k=1}^{n} \frac{p \mp i w_k(0)}{p \pm iw_k(0)} ~ r_{l,r}^{(0)}\label{e4}
\end{equation}
For a given complex crystal (whose reflection and transmission coefficients are known in advance) the reflection and transmission amplitudes for its isospectral crystals can be evaluated by using the relationship (\ref{e4}). However, for practical computation it is better to express $r_{l,r}^{(n)}$ in terms of the quantities associated with the initial crystal only. To do so, we use equations (\ref{e7}) and (\ref{e6a}) in equation (\ref{e4}) to obtain 
\begin{equation}
r^{(n)}_{l,r} = (-1)^n \textstyle \prod\limits_{k=1}^{n}\frac{p \mp i[\mathcal{W}_k(0)-\mathcal{W}_{k-1}(0)]}{p \pm i[\mathcal{W}_k(0)-\mathcal{W}_{k-1}(0)]} ~ r^{(0)}_{l,r}, \label{e11}
\end{equation}
where $\mathcal{W}_0(0) = 0$. A few remarks at this point are very important. For a $\mathcal{PT}$-symmetric complex potential, $\mathcal{W}_{(k)}(0)-\mathcal{W}_{(k-1)}(0)$ is complex valued. Therefore in contrast to the scattering by two real SUSY partners, complex $\mathcal{PT}$-symmetric isospectral potentials have $|r_{l,r}^{(n)}| \ne |r_{l,r}^{(0)}|$. Depending on the nature of the factorization functions at $x=0$, one has either $|r_l^{(n)}| < |r_l^{(0)}|$ and $|r_r^{(n)}| > |r_r^{(0)}|$ or vice versa. The equality of the transmission coefficients is accounted for by the equal asymptotic behavior of the potentials. The results obtained in equations (\ref{e4}) and (\ref{e11}) are consistent with the fact that the reflectances $\mathbb{R}_{l,r}^{(n)}$ and transmittance $\mathbb{T}^{(n)}$ of the n-th order isospectral crystal $V^{(n)}$ satisfy the generalized unitarity relation (\ref{e102}) (provided that the same quantities of the initial crystal $V^{(0)}$ do so).

\begin{figure*}[htb]
  \includegraphics[width=6.45cm,height=4cm]{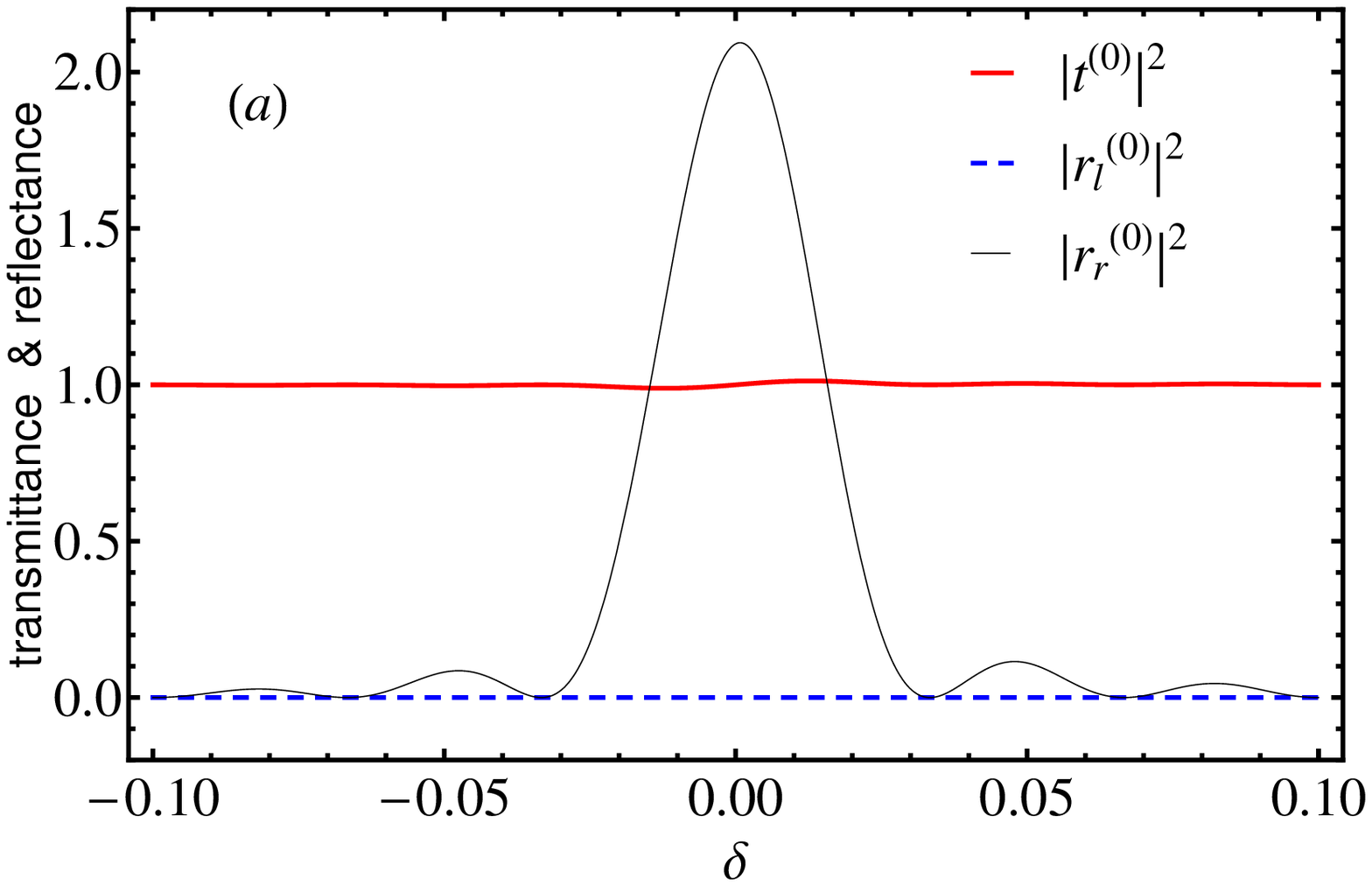} ~~ \includegraphics[width=6.45cm,height=4cm]{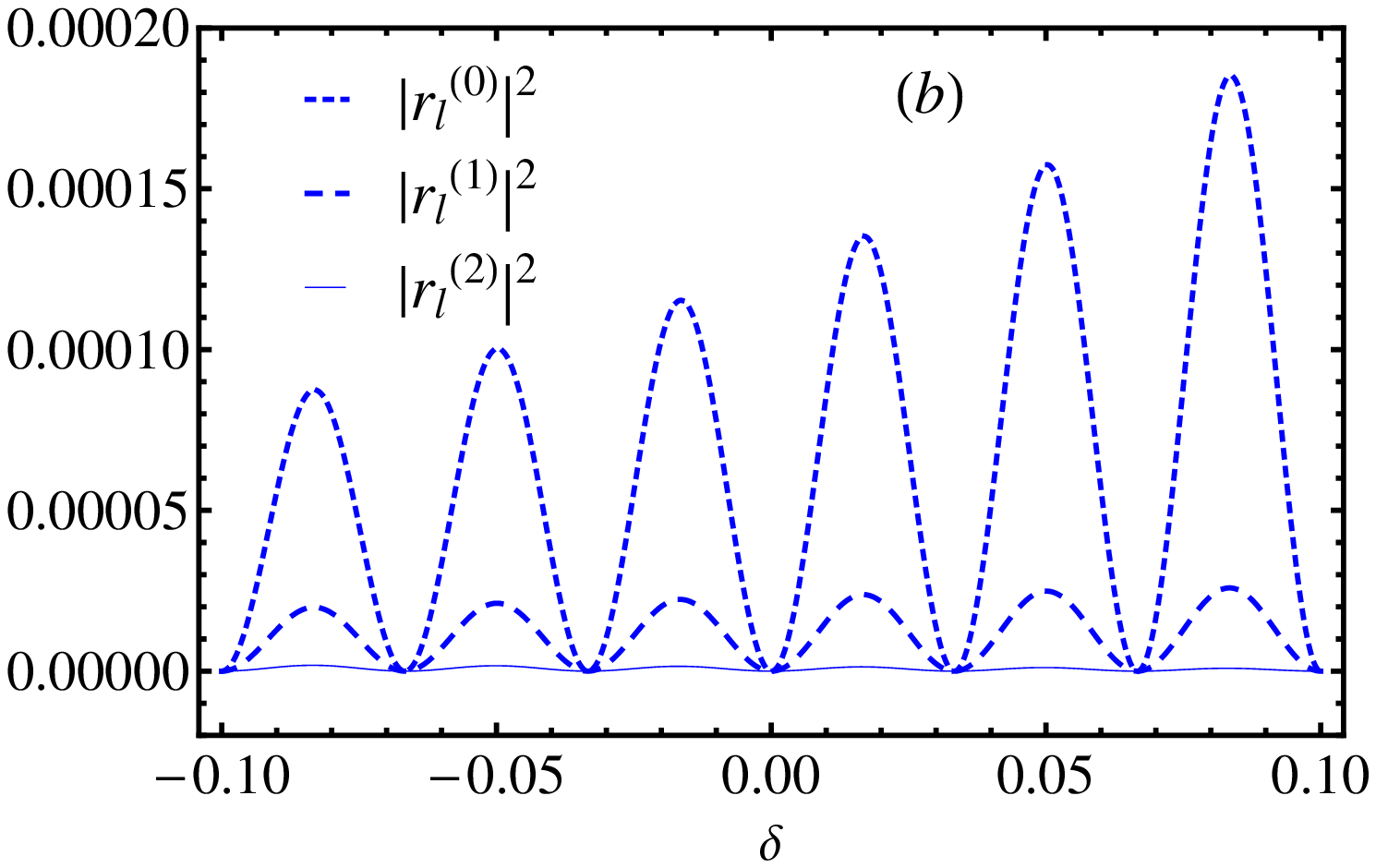} \includegraphics[width=6.4cm,height=4cm]{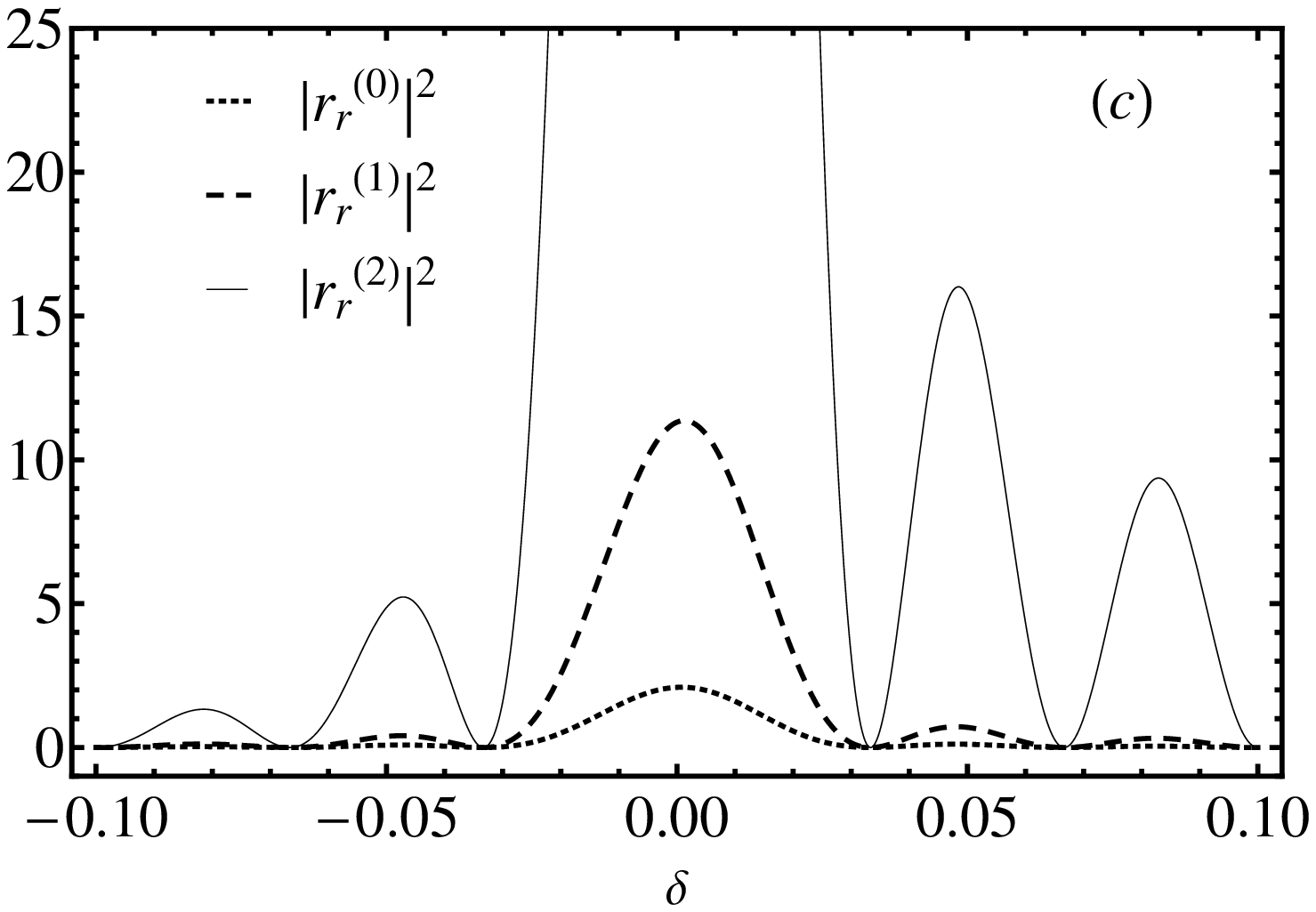} ~ \includegraphics[width=6.45cm,height=4cm]{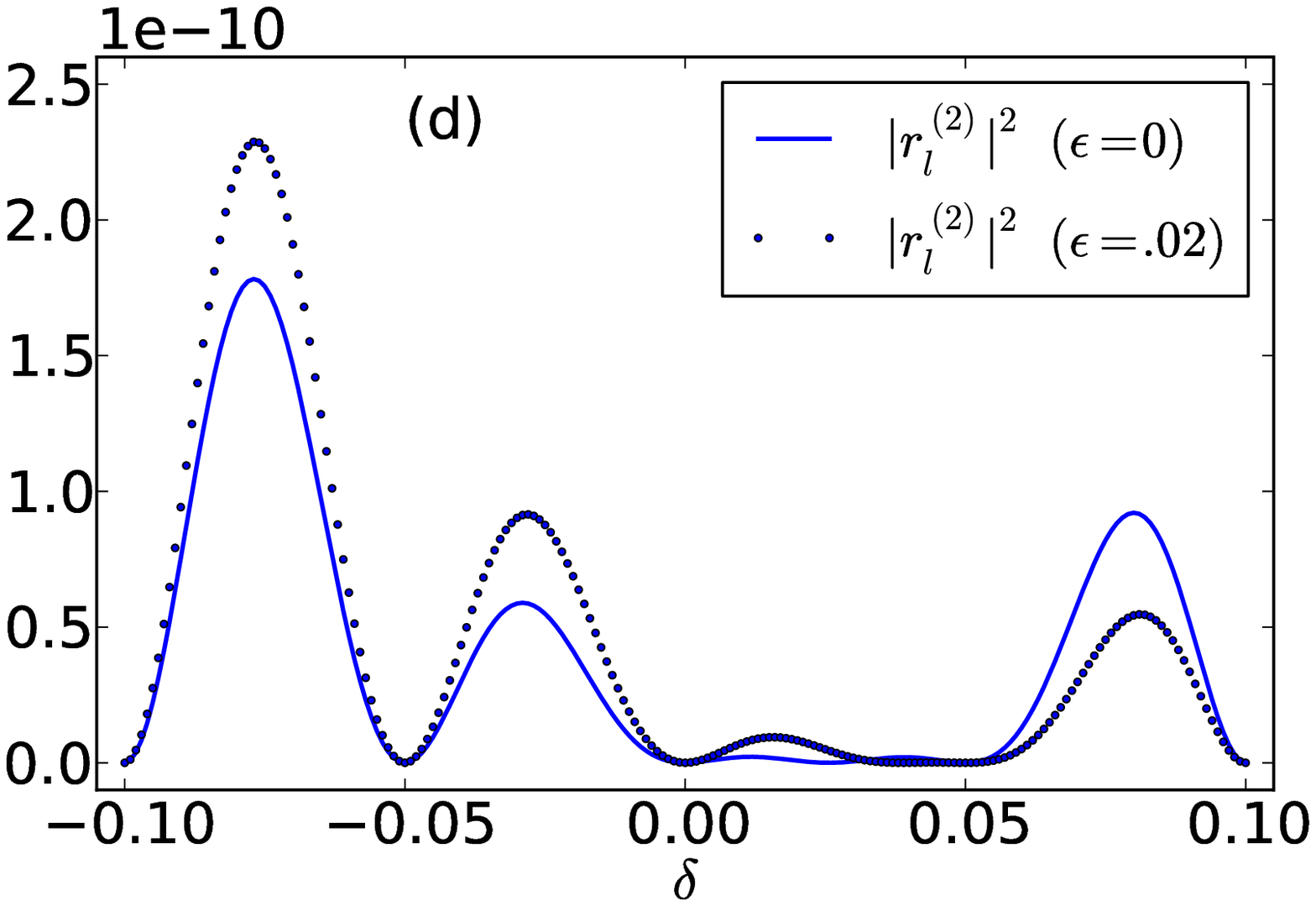}
   \caption{(Color online) (a) Transmittance $|t^{(0)}|^2$ and left, right reflectances $|r_{l,r}^{(0)}|^2$ for the crystal $V^{(0)}$ of length $L=30\pi$; comparison among (b) the left reflectances $|r_l^{(0),(1),(2)}|^2$ and (c) the right reflectances $|r_r^{(0),(1),(2)}|^2$ for the three crystals $V^{(0),(1),(2)}$ shown earlier in figure \ref{f1}. We considered the same parameter values of figure \ref{f1} to draw these three figures. (d) Solid curve is the magnified plot of the left-reflectance for the isospectral crystal  $V^{(2)}$ for $b=.001, \xi_0 = .01, \xi_1 = .95, L=20 a$ and $a = \pi$. The dotted curve represents the plot of the numerically computed left-reflectance for the perturbed potential $V^{(2)}+\epsilon \Delta V$ with $\epsilon =0.02$.}\label{f2}
   \end{figure*}

It is worth mentioning here that the reflection and transmission co-efficients for the crystal (\ref{e20}) are calculated analytically in ref.\cite{Lo11,Jo12}. In the present notations they are quoted in \cite{RT}. Using these exact expressions in equation (\ref{e11}) one can easily finds the explicit expressions for the scattering coefficients for the isospectral crystals. In figure \ref{f2}, we have shown the differences among the scattering co-efficients for the three crystals $V^{(0),(1),(2)}$ which are plotted earlier in figure \ref{f1}. Note here that the usual Bragg scattering condition occurs at the Bragg point $\delta = p - \pi/a$. Hence in the figure \ref{f2}, we have plotted (left, right) reflectance and transmittance with respect to the detuning parameter $\delta$. From the figure \ref{f2}(b) it is clear that the left reflectivity can be reduced drastically close to zero using higher order SUSY transformations. At the same time, figure \ref{f2}(c) shows that the corresponding right reflectivity can be enhanced. In figure \ref{f2}(d) we have shown the magnified picture of left-reflectance obtained after two SUSY transformations for $b= 0.001$. Clearly the left reflectance is much more close to zero compared to the one reported in \cite{Li+11}.

\FloatBarrier

From the experimental point of view, it is important to check the robustness of the predicted scattering behavior of $V^{(2)}$. To do so, we have numerically computed the scattering coefficients for the perturbed potential $V^{(2)} + \epsilon \Delta V$, where $\Delta V = e^{i z}$, $\epsilon \ll 1$ so that $|\epsilon \Delta V| \le \epsilon$. To obtain the scattering co-efficients associated to this perturbed potential, we have first evaluated the solution of the Schr\"odinger equation (\ref{e1}) in the interval $(0,L)$ using the 4th order Runge-Kutta method. The obtained solution and its first-order derivative have been then matched with those of the left and right propagating waves at the boundary $z=0$ and $z=L$, respectively. The reflectances and transmittance, so obtained, are found to be very close to the analytical results of the unperturbed potential. In particular, we have shown the left-reflectance for $\epsilon = 0.02$ in figure \ref{f2}(d) [dotted curve].\\

 In conclusion, we have shown that by a suitable extension of the SUSY method it is possible to construct transparent and one-way reflectionless crystals with sophisticated shape and structure. We hope that the present theoretical study would be a promising step towards the designing of a scatterer having a more pronounced invisibility effect. \\
 \\

   \section*{Acknowledgment}
   \noindent The author is a beneficiary of a postdoctoral grant from the Belgian Federal Science Policy
   Office (``BriX" IAP program P7/12) co-funded by the Marie Curie Actions from the European Commission.\\\\


\begin{thebibliography}{99}
   \bibitem{Le06} U. Leonhardt, Science, {\bf 312}, 1777 (2006).
   \bibitem{PSS06} J.B. Pendry, D. Schurig, and D. R. Smith, Science {\bf 312}, 1780 (2006).
   \bibitem{CCS10} H. Chen, C. T. Chan and  P. Sheng, Nature Materials {\bf 9}, 387 (2010).
      \bibitem{Va+09} J. Valentine, J. Li, T. Zentgraf, G. Bartal, and X. Zhang, Nature Materials {\bf 8}, 568 (2009).
      \bibitem{AE05} A. Alu, and N. Engheta, Phys. Rev. E {\bf 72}, 016623 (2005). 
      \bibitem{Sc06} D. Schurig et al., Science {\bf 314}, 977 (2006). 
    \bibitem{Zh+08} S. Zhang, D.A. Genov, C. Sun, and X. Zhang, Phys. Rev. Lett. {\bf 100}, 123002 (2008).
      \bibitem{FA13} R. Fleury and A. Alu, Phys. Rev. B {\bf 87}, 045423 (2013).   
 \bibitem{FEGM08}  M. Farhat, S. Enoch, S. Guenneau, and A. Movchan, Phys. Rev. Lett. {\bf 101}, 134501 (2008).
  \bibitem{ZXF11}  S. Zhang, C. Xia, and N. Fang, Phys. Rev. Lett. {\bf 106}, 024301 (2011).
  \bibitem{Ma+08} K.G. Makris, R. El-Ganainy, D.N. Christodoulides, and Z.H. Musslimani, Phys. Rev. Lett. {\bf 100}, 103904 (2008).
   \bibitem{Ru+10} C.E. Ruter, K.G. Makris, R. El-Ganainy, D.N. Christodoulides, M. Segev, and D. Kip, Nat. Phys. {\bf 6}, 192 (2010).
   \bibitem{Be08} M.V. Berry, J. Phys. A {\bf 41}, 244007 (2008). 
  \bibitem{Zh+10} M.C. Zheng, D.N. Christodoulides, R. Fleischmann, and T. Kottos, Phys. Rev. A {\bf 82}, 010103(R) (2010).
   \bibitem{Gu+09} A. Guo et al., Phys. Rev. Lett. {\bf 103}, 093902 (2009).
     \bibitem{Lo10R} S. Longhi, Phys. Rev. A {\bf 82}, 031801(R) (2010).
   \bibitem{CGS11} Y. D. Chong, Li Ge, and A. D. Stone, Phys. Rev. Lett. {\bf 106}, 093902 (2011).
   \bibitem{Sc10}H. Schomerus, Phys. Rev. Lett. {\bf 104}, 233601 (2010). 
   \bibitem{Ca+13} G. Castaldi, S. Savoia, V. Galdi, A Alu, and N. Engheta, Phys. Rev. Lett. {\bf 110}, 173901 (2013).
   \bibitem{Zh+13} X. Zhu, L. Feng, P. Zhang, X. Yin, and X. Zhang, Opt. Letters {\bf 38}, 2821 (2013).
     \bibitem{Li+11} Z. Lin, H. Ramezani, T. Eichelkraut, T. Kottos, H. Cao and D. N. Christodoulides, Phys. Rev. Lett. {\bf 106}, 213901 (2011).
     \bibitem{Lo11} S. Longhi, J. Phys. A {\bf 44}, 485302 (2011).
      \bibitem{Jo12} H. F. Jones, J. Phys. A {\bf 45}, 135306 (2012).
    \bibitem{Re+12} A. Regensburger, C. Bersch, M.A. Miri, G. Onishchukov, D.N. Christodoulides, and U. Peschel, Nature {\bf 488}, 167 (2012).
      \bibitem{YZ13} L. Feng et al., Nature Materials {\bf 12}, 108 (2013).
\bibitem{CW94} S.M. Chumakov and K. B. Wolf, Phys. Lett. A {\bf 193}, 51 (1994).
\bibitem{Mi+13}M. A. Miri, M. Heinrich, R. El-Ganainy, and D.N. Christodoulides, Phys. Rev. Lett. {\bf 110}, 233902 (2013).
   \bibitem{MHC13} M. A. Miri, M. Heinrich and D. N. Christodoulides, Phys. Rev. A {\bf 87}, 043819 (2013).
   \bibitem{LV13} S. Longhi and G. Della Valle, Ann. Phys. {\bf 334}, 35 (2013).
   \bibitem{LD13} S. Longhi and G. Della Valle, EurPhys. Lett. {\bf 102}, 40008 (2013).
      \bibitem{BC06} J. Bai and D. S. Citrin, Opt. Express {\bf 14}, 4043 (2006).
   \bibitem{Fe14} A. Zuniga-Segundo, B.M. Rodriguez-Lara, D.J. Fernandez, and H.M. Moya-Cessa, Opt. Express {\bf 22}, 987 (2014).
   \bibitem{Mo09} A. Mostafazadeh, Phys. Rev. Lett. {\bf 102}, 220402 (2009).
  \bibitem{Mo13} A. Mostafazadeh, Phys. Rev. A {\bf 87}, 012103 (2013).
   \bibitem{GCS12} L. Ge, Y.D. Chong and A. D. Stone, Phys. Rev. A {\bf 85}, 023802 (2012). 
   \bibitem{Ga80} M. G. Gasymov, Func. Ana. Appl. {\bf 14}, 11 (1980).
   \bibitem{CJT98} F. Cannata, G. Junker, and J. Trost, Phys. Lett. A {\bf 246} 219, (1998).
   \bibitem{MRR10} B. Midya, B. Roy and R. Roychoudhury, Phys. Lett. A {\bf 374}, 2605 (2010).
   \bibitem{GJ11} E.M. Graefe, and H.F. Jones, Phys. Rev. A {\bf 84}, 013818 (2011)
   \bibitem{CKS02} F. Cooper, A. Khare and U. Sukhatme, Supersymmetry in Quantum Mechanics, (World Scientific (2002)).
   \bibitem{DF98} G. Dunne, and J. Feinberg,  Phys. Re. D {\bf 57}, 1271 (1998). 
   \bibitem{Cr55} M. N. Crum, Quarterly J. Math. {\bf 6}, 121 (1955).
     \bibitem{Kr57} M. G. Krein, Dokl. Akad. Nauk SSSR {\bf 113}, 970 (1957).
     \bibitem{CDV07}F. Cannata, J.P. Dedonder, A. Ventura, Ann. Phys. {\bf 322}, 397 (2007).
   \bibitem{RT} The exact reflection and transmission coefficients for the crystal $V^{(0)}(z) = b \exp{2 i \pi z/a}$, as obtained in ref.\cite{Lo11}, are given by\\
    $t^{(0)} = \frac{2p\sin(\pi q)}{2p \sin(\pi q)\cos(pL)-a i \sin(pL)[p^2 I_q I_{-q}-b I'_q I'_{-q}]}$,\\
   
 $r_{l,r}^{(0)} = -\frac{a i \sin(pL)}{2p \sin(\pi q)} \left[b I'_q I'_{-q} + p^2 I_q I_{-q}\right.\\ ~~~~~~~~~~~~~~~~~~~~~~~~~~~ \left.\mp p\sqrt{b} (I'_q I_{-q} + I'_{-q} I_q)\right] t^{(0)}$,\\
  where $I_q = I_q(\Delta)$, $I'_{q} = I_{q-1}(\Delta) - (q/\Delta) I_{q}(\Delta)$, and $\Delta = a\sqrt{b}/\pi$, $q = ap/\pi$.
   
   \end{thebibliography}
 \end{document}